\documentclass[namedreferences]{solarphysics}
\usepackage[hyperref,optionalrh,solaromanenum]{spr-sola-addons} 
\usepackage{graphicx}                    
\usepackage{color}                       
\usepackage{breakurl}                         

\chardef\us=`\_

\begin{document}

\begin{article}

\begin{opening}

\title{Motion magnification in coronal seismology}

%
\author[addressref={warwick},corref,email={}]{\inits{S.A.}\fnm{Sergey}~\lnm{Anfinogentov}}
\author[addressref={warwick},corref,email={}]{\inits{V.M.}\fnm{Valery~M.}~\lnm{Nakariakov}}

%
\runningauthor{Sergey Anfinogentov and Valery M. Nakariakov}
\runningtitle{Motion magnification in coronal seismology}

\address[id={warwick}]{Centre for Fusion, Space and Astrophysics, Department of Physics, University of Warwick, CV4 7AL, UK}

\begin{abstract}
We introduce a new method for the investigation of low-amplitude transverse oscillations
of solar plasma non-uniformities, such as coronal loops, individual strands in coronal arcades, jets, prominence fibrils, polar plumes, and other contrast features, observed with imaging instruments. The method is based on the two-dimensional dual tree complex wavelet transform (DT$\mathbb{C}$WT). It allows us to magnify transverse, in the plane-of-the-sky,  quasi-periodic motions of contrast features in image sequences. The tests performed on the artificial data cubes imitating exponentially decaying, multi-periodic and frequency-modulated kink oscillations of coronal loops showed the effectiveness, reliability and robustness of this technique. The algorithm was found to give linear scaling of the magnified amplitudes with the original amplitudes provided they are sufficiently small. Also, the magnification is independent of the oscillation period in a broad range of the periods. The application of this technique to SDO/AIA EUV data cubes of a non-flaring active region allowed for the improved detection of low-amplitude decay-less oscillations in the majority of loops.
\end{abstract}

%

\end{opening}

%

\section{Introduction}

Diagnostics of physical parameters of solar coronal plasma structures with magnetohydrodynamic (MHD) waves, MHD seismology of the corona, is a popular and rapidly developing research technique \citep[see, e.g.][and references therein]{2014SoPh..289.3233L,2016SSRv..tmp....2N}. However, application of this method is limited by time and spatial resolution of available observational facilities: often the oscillations are detected near the very threshold of the resolution. Advanced data analysis techniques that could maximise the seismological outcome of the available data are then intensively developed in this context \citep[e.g.][]{2004ApJ...614..435T,2008SoPh..248..395S,2011A&A...526A..58S, 2012SoPh..275...79M, 2014ApJ...790L...2T, 2015A&A...574A..53K}. 

One of the important examples of coronal oscillations detected near the thres\-hold of currently available spatial resolution is the decay-less regime of kink oscillations, observed for the first time by \cite{2012ApJ...751L..27W}.  \cite{2013A&A...552A..57N} established that decaying and decay-less kink oscillations could coexist in the same loop. They investigated decaying oscillations triggered by an eruptive event  and found  decay-less oscillations with much lower displacement amplitude and the same period, occurring before and after the large-amplitude event. Decay-less oscillations were also observed in non-flaring active regions \citep{2013A&A...560A.107A, 2014A&A...570A..84N}, without any association with an eruptive event. Further investigation \citep{2015A&A...583A.136A} revealed that low amplitude kink oscillations are present in almost any active region (19 of 21 analysed). It was found that these oscillations most probably are standing kink waves, since the oscillation period linearly scales with the length of the loop. The constant phase distribution of detected kink oscillations along the oscillating loop also supports this interpretation. As decay-less kink oscillations are a common phenomenon in the solar corona and are seismologically sensitive to the magnetic field and plasma density, they  are a promising tool for coronal seismology.
However, the average amplitude of these oscillations was found to be about 0.2 arc-second \citep{2015A&A...583A.136A}, which is smaller than the pixel size of EUV imagers traditionally used for their detection, e.g. TRACE and SDO/AIA.

Likewise, an important parameter of another common class of kink oscillations of coronal loops, large-amplitude rapidly decaying oscillations \citep{1999ApJ...520..880A, 1999Sci...285..862N}, is the decay time that gives us information about transverse structuring of the plasma \citep[e.g.][]{2002A&A...394L..39G}. More recently, the attention was attracted to the existence of two regimes of the kink oscillation damping, exponential and Gaussian \citep{2013A&A...551A..39H, 2013A&A...551A..40P}. Discrimination between these two regimes can also provide us with seismological information, provided high-precision observations are available \citep{2016A&A...585L...6P}. In addition, simultaneous detection of several oscillation modes, e.g. different parallel harmonics, gives us important seismological information too, e.g. allowing for the estimation of the effective stratification and the magnetic flux tube expansion \citep{2005ApJ...624L..57A, 2008A&A...486.1015V, 2009SSRv..149....3A}. 
However, the application of this seismological technique is limited by the low amplitudes of higher harmonics \citep[e.g.][]{2004SoPh..223...77V, 2007A&A...473..959V, 2009A&A...508.1485V}. Interesting seismological information can also be obtained from the variation of the oscillation amplitude along the loop 
\citep{2007A&A...475..341V}, but its measurement also requires resolving low amplitude displacements.

Therefore, analysing such a low-amplitude oscillation is an important and challenging task. In this paper we present  a possible way to maximise the outcome of this analysis --- a method of the magnification of small transverse motions in time sequences of coronal imaging data.
 
\subsection{Motion magnification methods}
Motion magnification acts like a microscope for low amplitude motions in image sequences, i.e. imaging data cubes or videos. It artificially amplifies small displacements making them detectable by eye or some automated technique. First approaches in motion magnification \citep{citeulike:7234004} were based on explicit estimation of the velocity fields and subsequent warping individual frames to match amplified displacements. However, the approaches based on the motion estimation techniques were found to be computationally heavy and able to produce artefacts.  

	\cite{emoRecognition:Wu_2012}  developed the Eulerian video magnification technique that eliminates the explicit computation of the velocity field. It applies spatial decomposition to the input video sequence, followed by temporal filtering. The resulting signal is then amplified to reveal small intensity variations. This approach not only amplifies intensity variations but also transverse motions. However, only small values of magnification coefficients could be achieved for transverse motions. The other drawback of this method is that the brightness noise is amplified together with the intensity variations and, therefore, the output becomes rather noisy.
	
	The Eulerian video magnification technique was recently improved by \cite{Wadhwa:2013:PVM:2461912.2461966}. Their phase based method decomposes an input image sequence into complex steerable pyramids \citep{citeulike:3723307, citeulike:3808781} which are a kind of a 2D wavelet-like spatial transform. Since the individual components of such a decomposition are complex values, one can amplify temporal phase variation keeping the initial absolute values of the components. Such an approach amplifies only transverse motion, while the intensity variations remain almost untouched. The main advantages of the phase based approach in comparison with the method of \cite{emoRecognition:Wu_2012} are larger magnification coefficients and better signal to noise ratio in the output.
	
	In this paper, we follow the general ideas proposed by \cite{Wadhwa:2013:PVM:2461912.2461966} but use the dual tree complex wavelet  transform (DT$\mathbb{C}$WT) \citep{2005ISPM...22..123S} instead of complex steerable pyramids. In the acro\-nym of this method, the complex number symbol $\mathbb{C}$ is used to highlight its difference from the frequently-used acronym CWT for the continuous wavelet transform. Perfect reconstruction, good shift invariance, and computational efficiency are important features of DT$\mathbb{C}$WT. In contrast to methods developed by  \cite{Wadhwa:2013:PVM:2461912.2461966} and \cite{emoRecognition:Wu_2012} that are supposed to magnify motions inside a narrow temporal frequency band, our algorithm is aimed to amplify rather broadband motions. It allows us to address multi-modal and non-stationary, e.g. decaying and modulated, oscillatory processes often detected in solar plasma structures.
	
	In Section~\ref{sec:algorithm}, we give detailed description of our DT$\mathbb{C}$WT-based motion magnification algorithm. In Sections \ref{sec:model} and \ref{sec:observation} we apply our method to an artificial data set and real observations of the solar corona in extreme ultraviolet (EUV), respectively. In Section~\ref{concl} we summarise our conclusions.

\section{Algorithm description}
\label{sec:algorithm}
The presented method is designed to work with SDO/AIA data, but can be applied to any time sequence of images, i.e. an imaging data cube.
It utilises two-dimensional  DT$\mathbb{C}$WT \citep{2005ISPM...22..123S, rich_wareham_2014_9862} that decomposes an image $I$ into a set of complex-valued high-pass images $H_s$ of different scales and a low-pass residual $L$. The scale level $s$ is an integer number growing towards the largest scales. Thus, $H_0$ contains the finest details ($s = 0$) of the image. 

In DTW$\mathbb{C}$T,  wavelets are represented as filter banks. The detailed explanation of the filters design used in DT$\mathbb{C}$WT can be found in \cite{KINGSBURY2001234}.
	The Python DTW$\mathbb{C}$T implementation \citep{rich_wareham_2014_9862} provides
several filters of different lengths. 
		The length of the filter controls the localisation of the wavelet in spatial and wave number domains and defines a corresponding trade-off.
		Shorter filters provide better spatial resolution and ability to distinguish different motions in neighbouring structures, while the benefit of the longer filters is the weaker distortion at high magnification factors.
	Since the coronal loops are very often observed apparently close to each other or even overlapped, we have selected filters with the shortest length among all options included in the library: Near-Symmetric 5/7 tap filters (\textit{near\_sym\_a}) for the first level of DTW$\mathbb{C}$T, and Quarter Sample Shift Orthogonal (Q-shift) 10/10 tap filters (\textit{qshift\_a})  for the other scales.

\begin{figure*}
 \includegraphics[width=0.540\textwidth]{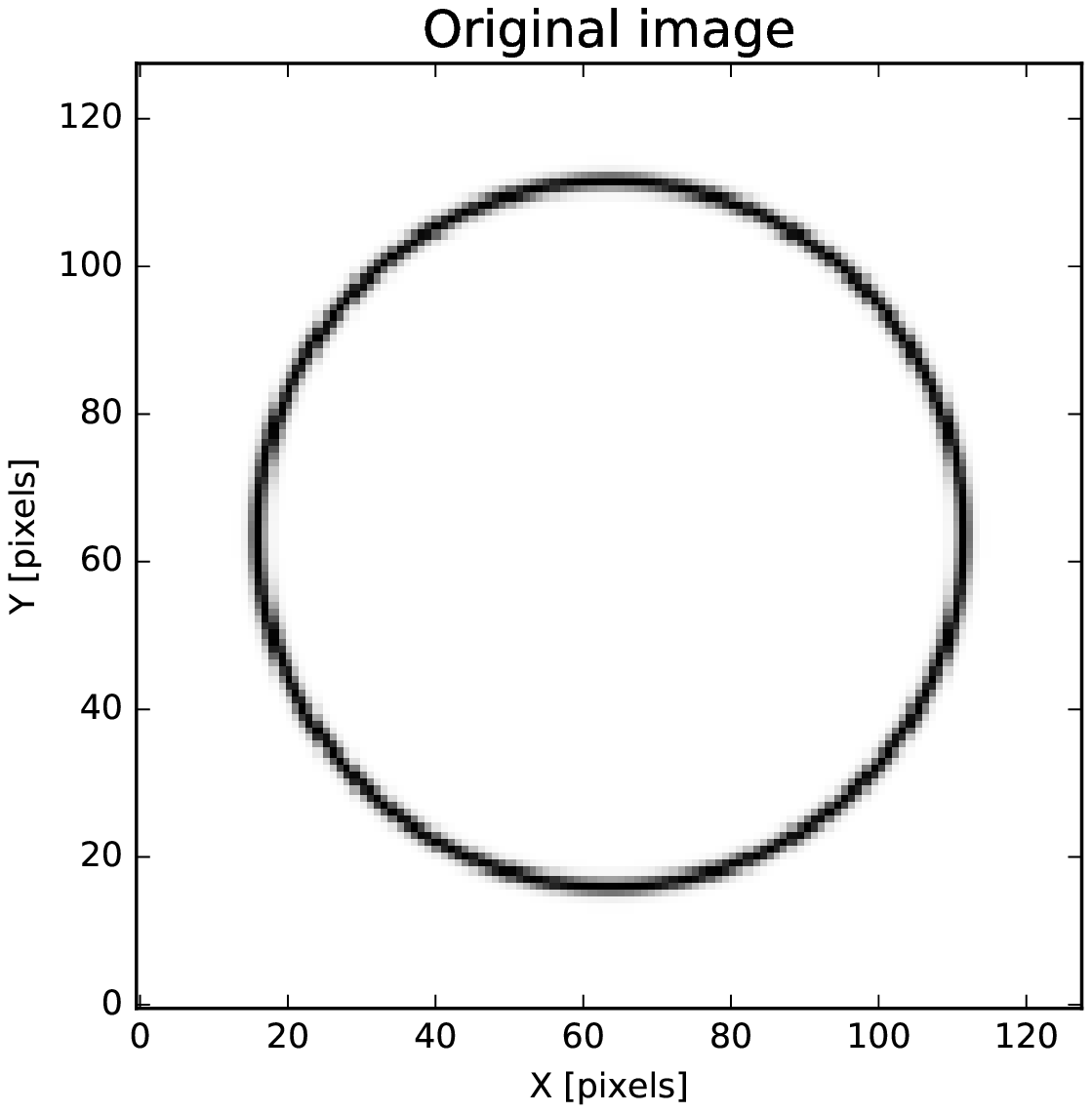}
 \includegraphics[width=0.410\textwidth]{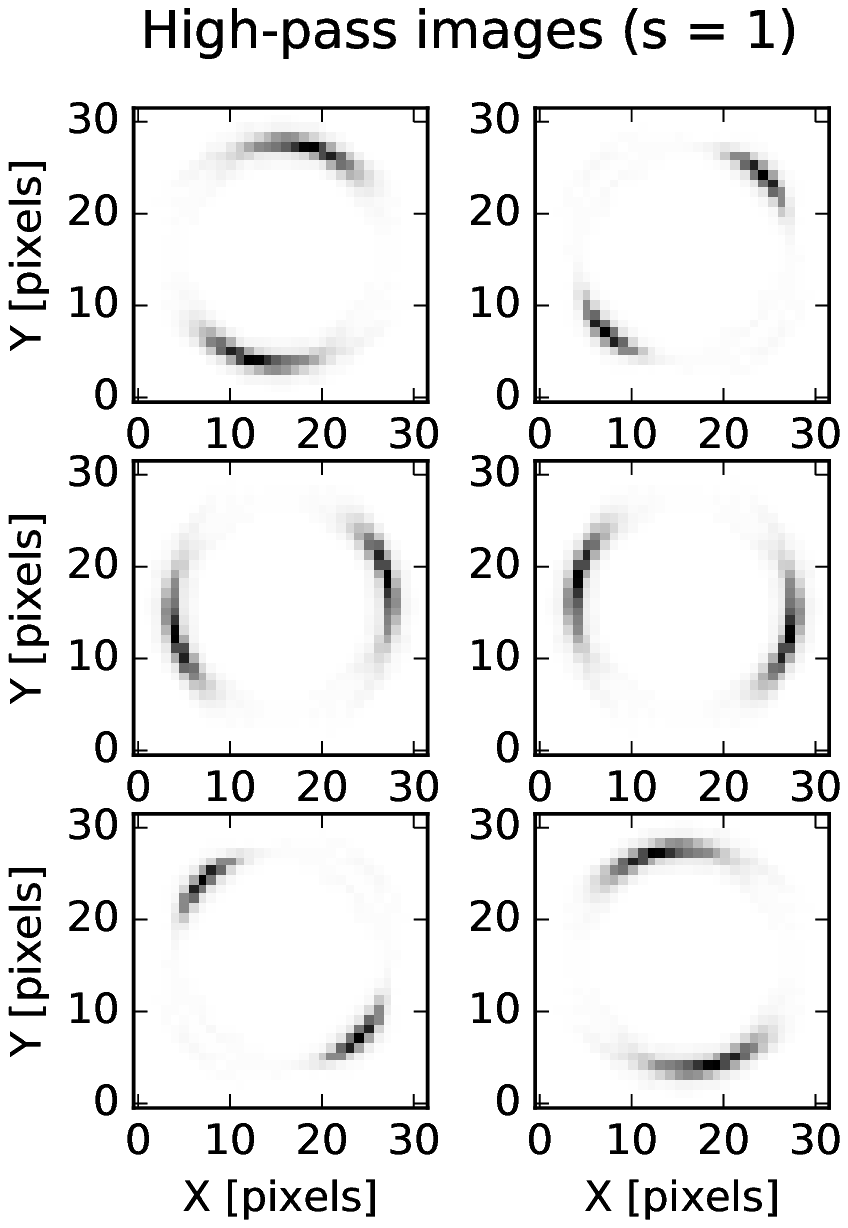} 
  \caption{Illustration of DT$\mathbb{C}$WT components of an annulus (\textit{left} panel). \textit{Right} panel shows absolute values of six high-pass images (scale level $s = 1$), corresponding to different wavelet orientations.  
}
\label{fig:orientations}
\end{figure*}

Unlike 1D wavelets, the basis of multidimensional wavelet transform is built not only by stretching and shifting the mother function but also by rotating it.
In a 2D case, DT$\mathbb{C}$WT uses six rotation angles, producing for each prescribed scale level $s$ six high-pass images, corresponding to different wavelet orientations.  Fig.~\ref{fig:orientations} gives an example of high-pass images of different orientations, obtained with the DT$\mathbb{C}$WT decomposition of an annulus (Fig. \ref{fig:orientations}, left).
The scale level $s =1$ is the same for all six images.
One can see that different segments of the annulus appear in different images. Decomposition of the image into several orientations allows us to process motions in different directions separately. 
In the following, we do not focus on the orientations for the purpose of simplicity, and notate this set of images as a single high-pass image $H_s$.

Let us consider two images $I^{t_1}$ and $I^{t_2}$ of a slowly moving object observed at different instants of time $t_1$ and $t_2$. 
For small displacements (i.e. less than 2--3 pixels), the absolute value of DT$\mathbb{C}$WT components and the low-pass residuals  remain almost the same ($H_s^{t_1} \approx H_s^{t_2}$, $L^{t_1} \approx L^{t_2}$).
The information about the object movement  is stored in the phases of high-pass images, $\phi(H_s^{t_1})$ and $\phi(H_s^{t_2})$. Here, $\phi(z)$ is the operation of getting the phase of a complex number $z$,  
\begin{equation}
\phi(z)  =  \Im[\ln(z)].
\label{eq:phase}
\end{equation}
One of the properties of DT$\mathbb{C}$WT is  that the phase difference $\Delta \phi = \phi(H_s^{t_2}) - \phi(H_s^{t_1}) $ depends linearly on the displacement of the object \citep{2005ISPM...22..123S}.
Hence, we can suppress or amplify these displacements in the data through manipulating with the phases of high-pass images.

The general algorithm of magnifying small motions in a data cube is the following: 
\begin{enumerate}
\item We compute spatial  DT$\mathbb{C}$WT of all frames taken in the time interval of interest. Each image $I^t$ is decomposed into a low-pass image $L^t$ and a number of complex valued high-pass images $H_s^t$,  corresponding to different spatial scale levels $s$. 
\item We compute phases of each pixel of high-pass images $H_s^t$. To eliminate phase discontinuities along the time axis, appearing due to the $2\pi$ periodicity,  we use the following formula for the phase extraction:
\begin{equation}
\Phi_s^t(x,y) = \sum_{\tau = 0}^t \phi(H_s^{\tau+1}(x,y)/H_s^{\tau}(x,y)).
\end{equation} 
\item Then we compute the phase trend $\overline{\Phi_s^t}$ by smoothing the phase $\Phi_s^t$ along the time axis by convolving it with a flattop window of the time width  $w$ on its half height.
The advantage of the flattop window is that its Fourier transform has a main lobe with a flat top and low level side lobes, reducing the impact of filtering on the amplitude of a filtered signal  in comparison with  the Gaussian and boxcar windows.  
\item The high frequency variations of the phase are calculated by subtracting the  trend from the phase, $\Phi_s^t - \overline{\Phi_s^t}$. The result  is then multiplied by a magnification factor $k>1$, amplifying all movements of the time-scales less than $w$.
\begin{equation}
(\Phi_{\mathrm{out}})_s^t = \overline{\Phi_s^t} + k(\Phi_s^t - \overline{\Phi_s^t}).
\end{equation} 
\item Unfortunately, the high-frequency noise in the phase is amplified together with the oscillatory signal. The noise can be reduced through optional smoothing of the output phase $\Phi_{\mathrm{out}}$ along the time axis. The smoothing is done by convolving the amplified phase with the flattop window of small size. To avoid appearance of the false oscillations the smoothing window must be shorter than the expected oscillation period. Our code uses the smoothing width of 2 time steps.
\item The output high-pass images are then made by combining the original absolute values and the output phase computed at the previous step,
\begin{equation}
	(H_{\mathrm{out}}) _s^t= |H_s^t|\, \exp{i(\Phi_{\mathrm{out}})_s^t}.
\end{equation} 
\item During the last step the output images are obtained  from the modified high-pass and original low-pass images using the inverse DT$\mathbb{C}$WT. 
\end{enumerate}

\section{Testing on artificial signals}
\label{sec:model}

\begin{figure*}
 \includegraphics[width=1\textwidth]{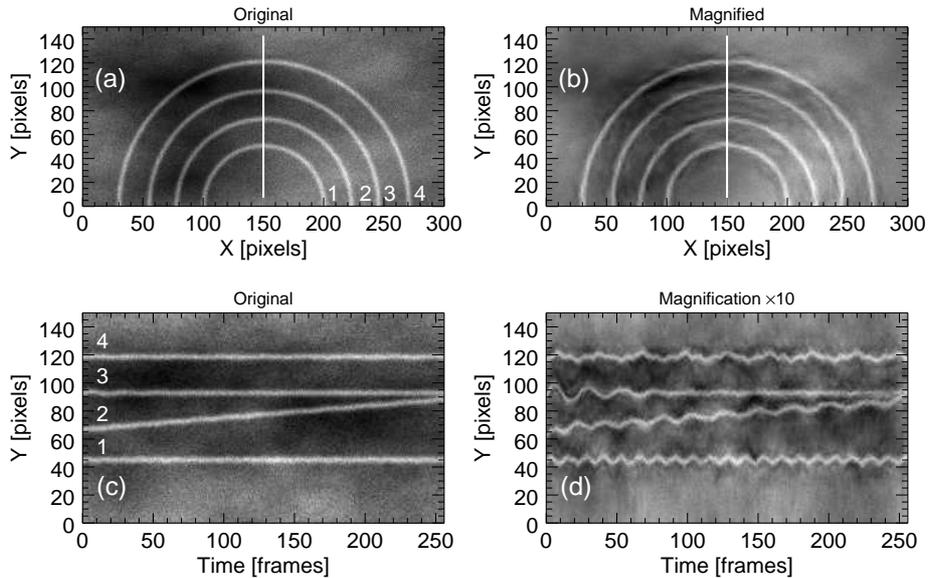} \caption{Detection of synthetic signals by the method of the DT$\mathbb{C}$WT-based motion magnification. Panel (a) shows a frame with four loops superimposed with the time varying spatially non-uniform background. Loop 1 oscillates harmonically with a constant period of 30 frames and amplitude 0.2~px. Loop 2 oscillates with constant amplitude of 0.2~px; its major radius increases with time, and the oscillation period increases with the increase in the loop length. Loop 3 performs a decaying oscillation with the initial amplitude of 1~px, and the damping time of 40 frames. Loop 4 performs a two-harmonic oscillations, with the period ratio of 3, the amplitude of the longer period oscillation is 0.2~px, and of the shorter period is 0.1~px. The white vertical line shows the location of the slit for time-distance maps.
Panel (b) shows the same frame as (a) but after the application of the DT$\mathbb{C}$WT-based motion magnification. Panel (c) gives the time-distance plot made along the slit shown in panel (a) for the original synthetic data. Panel (d) shows the time-distance plot after the application of the DT$\mathbb{C}$WT-based motion magnification.
}
\label{fig:model}
\end{figure*}

The method has been tested on a sequence of artificial images, imitating typical kinds of kink oscillations in a system of four loops superposed with the active background that is slowly varying in time and in space(see Fig.~\ref{fig:model}).
The images are taken at a constant time rate. The Poisson noise was applied to the  synthetic image sequence to imitate the photon noise on CCD matrix of e.g. SDO/AIA.

Loop 1 oscillates harmonically with a constant period of 30 time frames (the time the images are taken) and constant amplitude of 0.2~pixels in the plane-of-the-sky.
Loop 2 supports a standing kink oscillation corresponding to the global spatial parallel harmonic, imitating a typical decay-less kink oscillation of a coronal loop. The displacement amplitude is constant, of 0.2~pixels. The period increases in time, reflecting the gradual increase in the loop's major radius, at the speed of 0.08~pixel per time frame.
Loop 3 performs an exponentially decaying standing oscillation corresponding to the global spatial parallel harmonic, with the initial amplitude of 1~pixel, and the damping time of 40 time frames --- a typical decaying kink oscillation of a coronal loop. The oscillation of Loop 4 consists of two harmonics. The period ratio of these two harmonics is 3, imitating the superposition of the global mode and the third spatial parallel harmonic. The amplitude of the long period oscillation is 0.2~pixels, and of the short period is 0.1~pixels.

Recent studies  of  dynamical
 processes in the solar atmosphere \citep{2014A&A...563A...8A,2015ApJ...798....1I,2016A&A...592A.153K} 
 show that the active background (with spatial non-uniformities of the brightness, varying in time) in EUV solar data  behaves like a correlated coloured noise, having a power-law dependence of the spectral energy upon the frequency 
within broad range of periods from minutes to tens of hours. 
 According to this finding, we model the evolving background as 3D (two spatial coordinates plus time) coloured noise by filtering it out from the white noise data set, by applying a power law shaped low-pass filter in the 3D Fourier domain. Results of such filtering visually  resemble an evolving cloud, and is a suitable model for  active background slowly varying both in space and in time.

The artificial data cube was processed with the DT$\mathbb{C}$WT-based motion magnification algorithm, with the amplification factor $k = 10$. The time width of the smoothing filter was selected to be equal to 35 frames, that is slightly longer than the longest oscillation period (30~frames) in the data set. In Fig.~\ref{fig:model} we show snapshots of the loop system, taken in the original synthetic data cube and in the data cube after processing according to the algorithm described in Sec.~\ref{sec:algorithm}. 

In the bottom panels of Fig.~\ref{fig:model} we show time-distance plots made along the vertical line passing through the tops of the loops, for both original and processed data cubes. The figure clearly demonstrates the effectiveness and reliability of the developed technique. Indeed, the built-in oscillations are clearly visible in the processed data for all four kinds of the kink oscillatory patterns that are difficult to see in the original data. Fig.~\ref{fig:model}~(c) and (d). In all the cases the sub-resolution oscillations become evident with the naked eye after the magnification. Moreover, the oscillations are clearly resolved even when two loops are very close to each other (the oscillation in Loop 2 at the time about 250 frames, see panel (d)). In the latter case we see in Loop 3 some traces of the oscillation in Loop 2, that should be considered as an artefact of the developed method. We may also point out some artificial diffusion of the loop cross-section in the very beginning and end of the data series. 

\subsection{Dependence on the oscillation parameters}

One of the main properties of any spatial wavelet decomposition is the localisation of its components both in the spatial and wave number domains. The scale of this localisation is prescribed by 
the size of the envelope of the wavelet mother function in the spatial domain or by the length of the finite impulse response of wavelet filters. 

The spatial localisation of a wavelet component helps us to distinguish between different periodicities in the spatially neighbouring structures, because every wavelet component \lq\lq feels\rq\rq\ only the intensity distribution in the vicinity of its own position.
Thus, a change of a certain wavelet component will modify the reconstructed image only in the local spatial neighbourhood of that particular component. Consequently, the spatial support of a wavelet implies a limitation onto the maximum possible amplitude of magnified motions.

This limitation is also evident in the right panel of Fig. \ref{fig:orientations} that shows the absolute value of the wavelet components at the scale level $s~=~1$. Since the magnification procedure modifies only the phase of the wavelet components, it can not translate the segments of the annulus presented in Fig.~\ref{fig:orientations}  to the areas where the absolute value of the corresponding wavelet components  is weak.  Thus, if the expected output amplitude of the structure motion is too high, an oscillating structure will be significantly distorted in the reconstructed images. 

\begin{figure}
\includegraphics[width=0.5\textwidth]{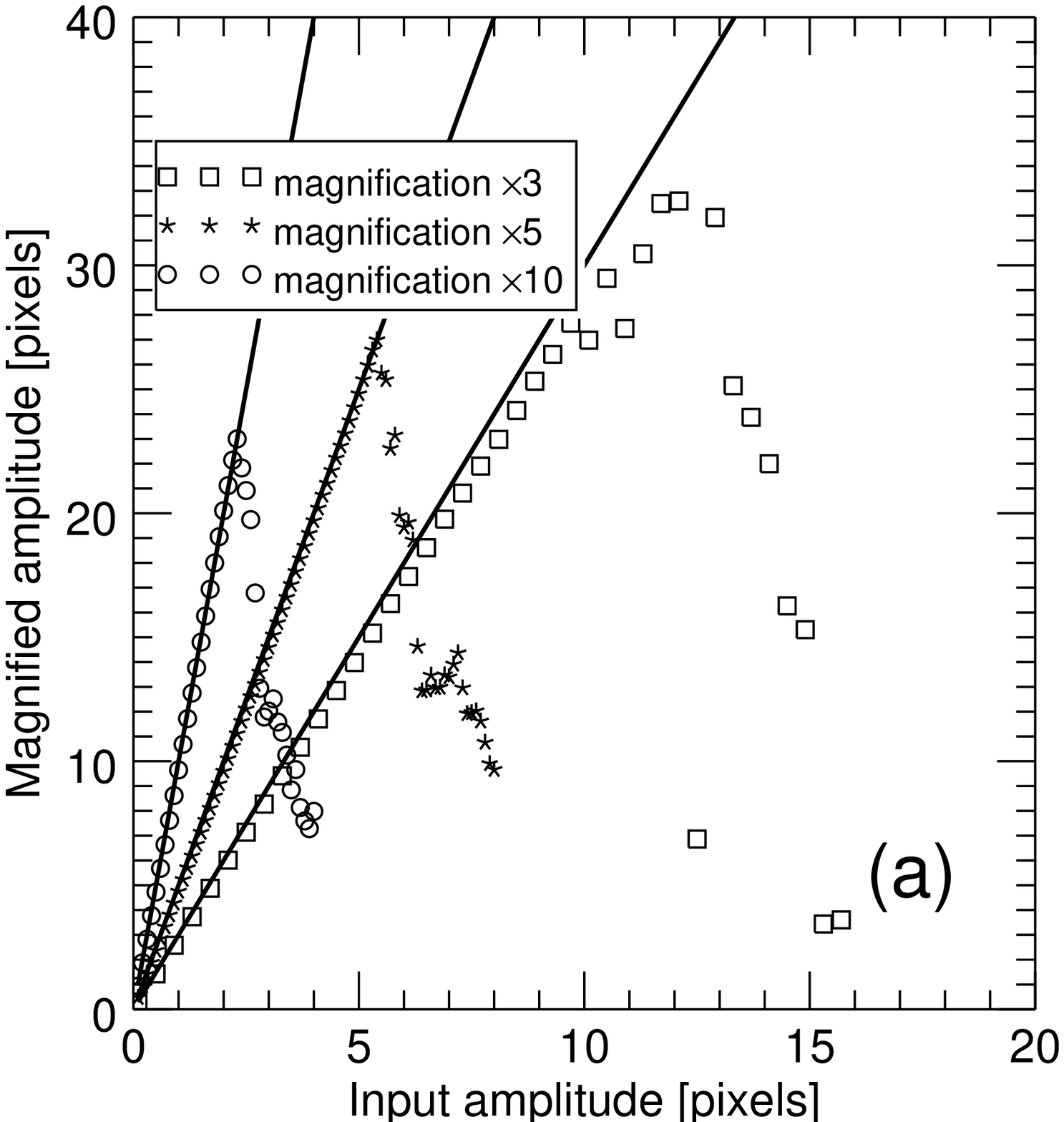}\includegraphics[width=0.5\textwidth]{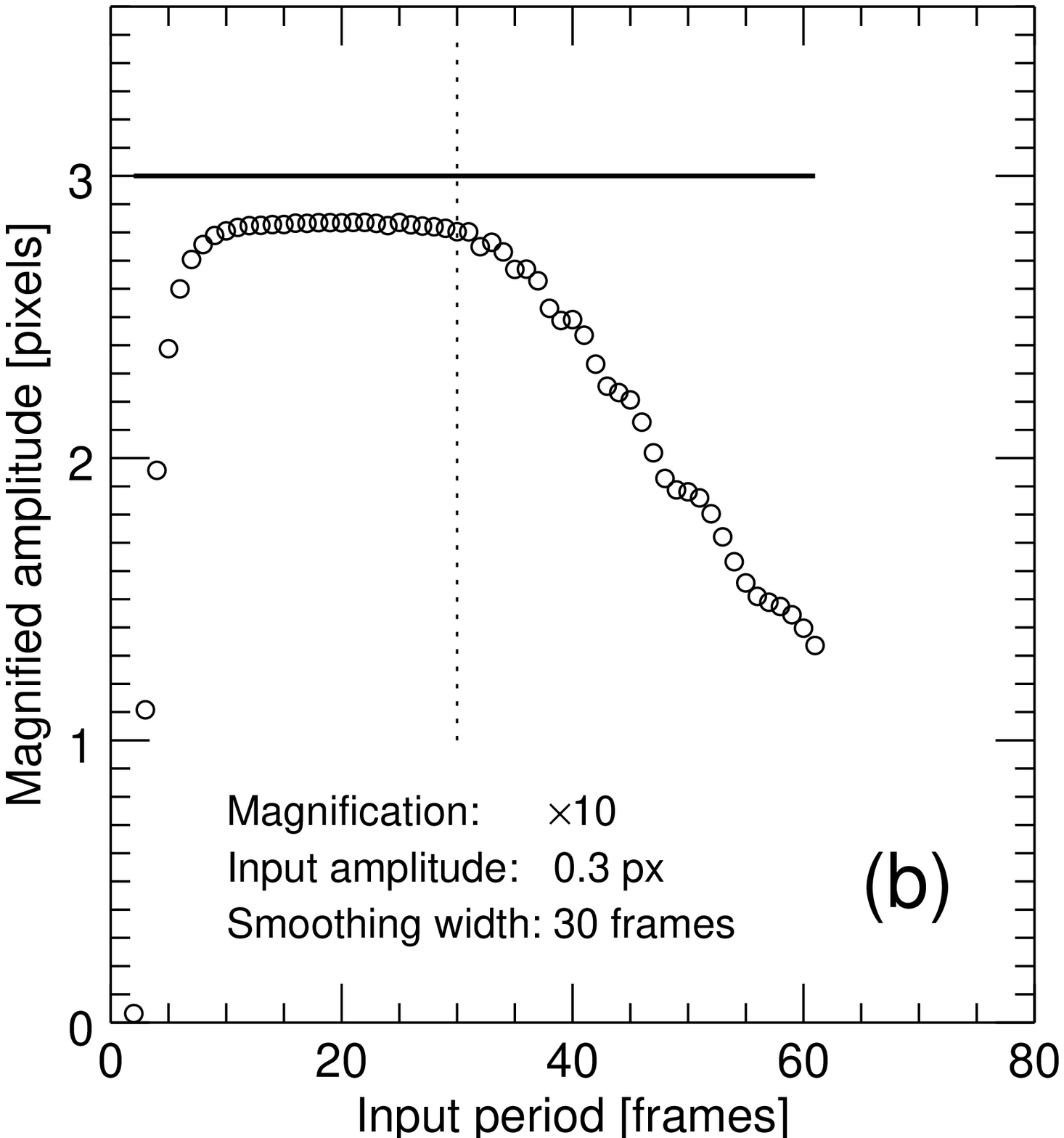}
 \caption{Behaviour of the DT$\mathbb{C}$WT-based motion magnification depending on the oscillation parameters. Panel (a) shows the dependence of the  magnified oscillation amplitude upon the input amplitude for the magnification factors $\mathbf{k = 3}$ \textbf{($\square$)}, $k = 5$ ($\ast$), and $k = 10$ ($\circ$). Panel (b) shows the dependence of the magnified amplitude upon the original oscillation period. The vertical dotted line indicates the size of the smoothing time window $w$ that was set to 30 time frames. The solid lines in both plots indicate expected ideal dependences.}
\label{fig:non-linear} 
\end{figure}

 Therefore, the magnified oscillation amplitude depends linearly upon the original one only in the low-amplitude range. To assess experimentally this range we processed a set of artificial data with different values of the initial oscillation amplitude. The scaling of the output amplitude with the input amplitude for the magnification coefficients $k=3$, $k=5$, and $k=10$ are presented in Fig.~\ref{fig:non-linear}~(a). 

The amplification factor $k =10$ gives linear scaling of the magnified and original amplitudes when the magnified amplitude is less than 25 pixels corresponding to the initial amplitude of 2.5 pixels (see Fig.~\ref{fig:non-linear}~(a)).
Similar  results are obtained for $k = 5$, and $k=5$. The magnification is saturated when the output amplitude is as large as 27 and 33 pixels, respectively, allowing us to magnify oscillations with the original  amplitude up to 5.4 pixels for $k=5$, and up to 10 pixels for $k =3$. Note, that transverse oscillations with the amplitude larger than 1--2 pixels are well seen in time-distance plots without any magnification, and hence do not require any magnification to be analysed anyway. We recommend to apply the developed technique to real data only when the magnified oscillation amplitude is expected to be less than 10 pixels. It guaranties that the magnification procedure is linear.

Since the transverse motions are magnified with respect to the low-frequency trend, the actual amplification coefficient depends upon the temporal frequency of the oscillation. 
Fig.~\ref{fig:non-linear}~(b) shows the dependence of the magnified  amplitude upon the input oscillation period. The figure demonstrates that 
the DT$\mathbb{C}$WT-based motion magnification is period-independent in the broad range of periods from 5--6 time frames up to the width $w$ of the smoothing time window. The latter is indicated in the right panel of Fig ~\ref{fig:non-linear} with a vertical dotted line. We recommend to set the smoothing width to be greater than the expected oscillation period, otherwise the oscillation will be magnified much less than expected.  A sharp increase in the magnification coefficient at the periods of about 4--6 time frames could lead to the appearance of false oscillations at those periods, filtered from the high-frequency noise. Thus, short-period (less than 10 time frames) oscillations found in the magnified data could be artefacts and must be analysed with caution.

\subsection{Dependence on the loop parameters}

Detection of coronal loop oscillations in EUV data is generally obstructed by two circumstances: non-uniform background emission varying in time,  and movements of the neighbouring coronal loops. Indeed, both of them can potentially lead to degradation of the motion magnification and appearance of pronounced artefacts in the output data cube.
In the following, we investigate the robustness of the motion magnification  with respect to the level of slowly evolving  background and distance to a neighbouring loop that oscillates with a different period.

We built a  simple model containing a straight bright structure  oscillating with a period of 30 time frames and amplitude of 0.3 pixels, superposed with an active background. The active background was modelled in the same way as   in Section~\ref{sec:model}.

\begin{figure*}
	\includegraphics[width=1.0\textwidth]{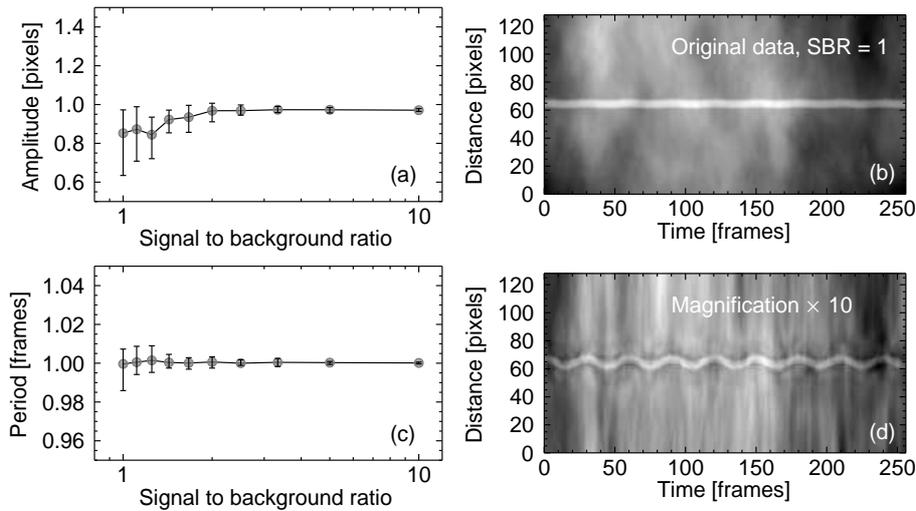} \caption{ Dependence of the amplitude (a) and period (c) inferred from the motion magnification results, upon the relative level of the background activity level.
The values of the amplitude and period are normalized to the known parameters of the model accounting for the magnification coefficient.
The parameters were independently estimated for 512 cross-sections extracted from the magnified data cube.
The grey circles in panels (a) and (c) show median values  of the obtained parameter distributions while the errors were estimated as 5\% and 95\% percentiles.
The right panels demonstrate the time distance plots of the original data (b) and motion magnification results (c).
The plots  correspond to the standard deviation of the background activity being equal to the  intensity of the oscillating structure.}
\label{fig:model-background}	 
\end{figure*}

We applied our method to magnify all transverse motions by a factor of 10.
Then we extracted 512 time-distance plots from the magnified data cube and estimated the oscillation amplitude and period with the standard method used in coronal seismology. First, we found the loop centre positions at each instance of time by fitting a Gaussian in the transverse intensity profile.
Then the evolution of the loop centre position was fitted with a sinusoid, providing us with the estimation of the main oscillation period and amplitude.
Thus, we got 512 independent measurements of the oscillation period and amplitude.
The detected values were estimated as sample medians (grey circles in Fig. \ref{fig:model-background}a and \ref{fig:model-background}c), while the confidence intervals  were defined as 5\% and 95\% percentiles of the  distributions.

The procedure described above was applied to the models with different signal to background ratios (SBR). Here, SBR means the ratio between the loop profile intensity maximum in the absence of the active background, and the standard deviation of the active background.
The time-distance plots extracted from the original data and from the motion magnification ($\times 10$) results corresponding to the  model with SBR = 1 are presented in Fig. \ref{fig:model-background}b and \ref{fig:model-background}d, respectively. 

Fig. \ref{fig:model-background}a and \ref{fig:model-background}c show that the  DT$\mathbb{C}$WT-based motion magnification satisfactorily works in the presence of  active background even when the background intensity variation amplitude is comparable to the relative intensity of the oscillating structure. However, higher variations of the active background lead to higher uncertainties  in the measured oscillation parameters, especially the amplitude (see Fig.~\ref{fig:model-background}).

	Coronal loops seen in EUV images of the Sun are often located very close to each other or even overlap.
	In this sense, the ability of the motion magnification method to resolve  different periodicities in neighbouring structures is crucial for the correct estimation of the oscillation parameters in the processed data.
	
	To assess the performance of   DT$\mathbb{C}$WT-based motion magnification in the case of neighbouring oscillating structures, we made  a model of two  loops situated close to each other. Both loops have a Gaussian intensity profile with the half-width of 2 pixels, and oscillate with the amplitude of 0.2 pixels. 
	The magnification factor was set to $\times 10$  giving the oscillation amplitude of two pixels in the processed data.
	
\begin{figure*}
	\includegraphics[width=1.0\textwidth]{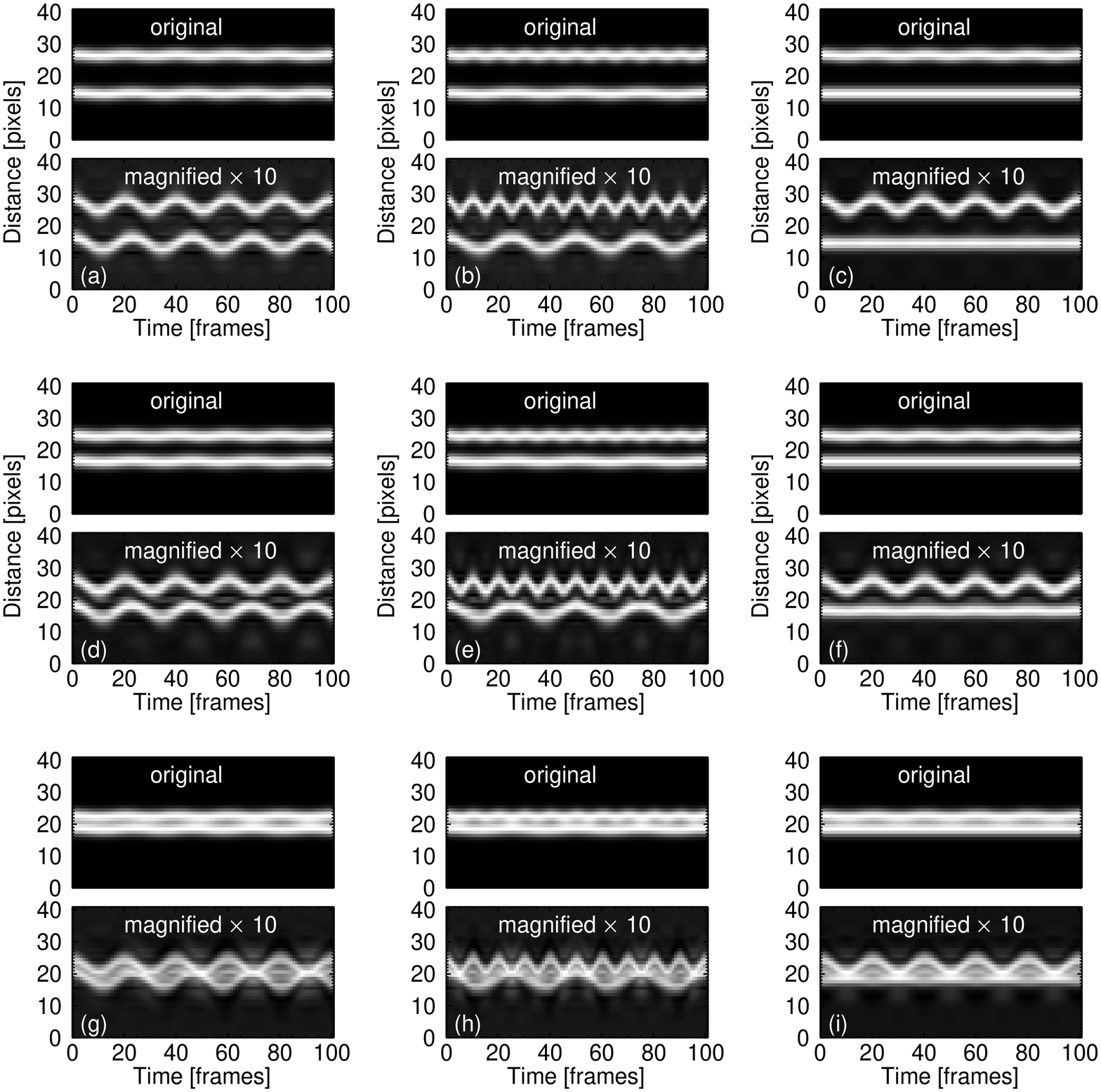} \caption{Ability of  the motion magnification to distinguish between different oscillation periods in neighbouring loops, depending on the distance between them.
Three different kinds of models are considered: two oscillating loops with slightly different periods of 20 and 23 time frames  (panels a,d, and g), two loops with significantly different periods of 10 and 25 frames (panels b,e, and h), and a steady loop in the vicinity of another one that oscillates with the period of 20 frames (panels c,f,i).
All loops have a Gaussian transverse profile with the half width of 2 pixels.
The oscillation amplitude is 0.2 pixels, giving the magnified ($\times 10$) amplitude of 2 pixels.
The distance between the loop centres is 12 pixels in panels (a) -- (c), 8 pixels in panels (d) -- (f), and 4 pixels in panels (g) -- (i).}
\label{fig:model-distance}	 
\end{figure*}

	The test results of the DT$\mathbb{C}$WT-based motion magnification applied to the model of two neighbouring loops are shown in Fig. \ref{fig:model-distance}. We considered three different oscillation patterns: 
	\begin{itemize}
		\item Two neighbouring loops  oscillate with slightly different periods of 20 and 23 time frames. The time-distance maps extracted from the original and processed data cubes are presented in  panels (a), (d), and (g) of Fig. \ref{fig:model-distance}.
		\item Two loops have significantly different oscillation periods (10 and 25 time frames). The corresponding time-distance plots are shown in panels (b), (e), and (h)of Fig. \ref{fig:model-distance}.
		\item One of the neighbouring loops is steady and the other one oscillates with the period of  20 frames. The motion magnification results are presented in panels (c), (f), and (i) of Fig. \ref{fig:model-distance}.
	\end{itemize}

We examined the performance of the DT$\mathbb{C}$WT-based motion magnification for various distances between the loops' cross-sectional centres:
\begin{itemize}
\item 12 pixels. The oscillating loops are located close to each other, but neither touch nor overlap, in both the  original and magnified data (panels (a -- c) of Fig. \ref{fig:model-distance}).
\item 8 pixels. In the magnified data, the oscillating loops  touch each other at the maximum displacement positions (panels (d -- f) of Fig.~\ref{fig:model-distance}).
\item 4 pixels. It is an extreme case when the loops apparently touch each other in the original data and became  overlapped after motion magnification (panels (g -- i) of Fig. \ref{fig:model-distance}).
\end{itemize}

Time-distance plots presented in Fig. \ref{fig:model-distance} demonstrate that DT$\mathbb{C}$WT-based motion magnification allows for distinguishing between different periodicities in neighbouring loops.
However, there could be an artificial  cross-influence between the oscillating loops. In particular, the oscillatory displacement pattern can slightly deflect from the original one, when the oscillating loops are sufficiently close to be overlapped in the magnified data. 
One can see this effect in Fig. \ref{fig:model-distance}e, by comparing the extreme positions of the lower loop at the times about 30 and 50 time frames.
Fortunately, this weak effect does not prevent the correct measurement of the oscillation amplitude and period.
Moreover, the oscillation parameters can be estimated for both loops even in the extreme case when these two structures become overlapped after the motion magnification (see Fig.~\ref{fig:model-distance}(g~--~i)).

We measured amplitudes and periods of transverse oscillations in magnified data using the same method as in all previous tests. The measurement results are presented in Table \ref{tab:distance}. The estimated values of the period are sufficiently close to the parameters set in the model in all analysed cases. However, the oscillation amplitude is slightly underestimated by about 10\% which is similar to the results of other tests presented in Fig. \ref{fig:non-linear} - \ref{fig:model-distance}, except the overlapping case where the measured amplitude appeared by about 5\% larger than the exact value. In the case when a steady loop is situated near the oscillating one, our measurements show the occurrence of an artificial oscillation in the steady loop with the same period as that of the neighbouring oscillating loop. Fortunately, the amplitude of this false oscillation is reduced by one order of magnitude even in the case of apparently overlapping loops. This artefact could be mitigated by checking whether there is a similar oscillatory pattern in another segment of the steady loop, which is situated  far from the neighbouring loop, if the loop geometry allows it.

\begin{table}
\caption{Parameters of the magnified oscillations, measured for two neighbouring loops, depending on the distance between them for the cases shown in Fig. \ref{fig:model-distance}. Three different kinds of models are considered: two oscillating loops with slightly different  periods, two loops with significantly different periods, and a steady loop in the vicinity of an oscillating one. In all cases, the coefficient of magnification is 10.}
\label{tab:distance}
\begin{tabular}{lcccc}
\hline
Measured&Exact&\multicolumn{3}{c}{Distance between loops}\\
parameter&value&12 px &8 px&4 px \\
\hline
\multicolumn{5}{c}{Slightly different periods (Fig. \ref{fig:model-distance}-adg)}\\
Period (top loop), frames&20&20.0$\pm$0.1&19.9$\pm$0.1&20.1$\pm$0.1\\
Amplitude (top loop), px&2&1.76$\pm$0.11&1.59$\pm$0.13&2.11+0.10\\
Period (bottom loop), frames&23&23.0$\pm$0.1&23.2$\pm$0.2&22.8$\pm$0.1\\
Amplitude (bottom loop), px&2&1.79$\pm$0.10&1.78$\pm$0.12&2.15$\pm$0.12\\
\multicolumn{5}{c}{Significantly different periods (Fig. \ref{fig:model-distance}-beh)}\\
Period (top loop), frames&10&10.0$\pm$0.02&10.0$\pm$0.04&10.0$\pm$0.02\\
Amplitude (top loop), px&2&1.75$\pm$0.10&1.60$\pm$0.14&2.09$\pm$0.10\\
Period (bottom loop), frames&25&25.0$\pm$0.2&25.0$\pm$0.2&25.0$\pm$0.2\\
Amplitude (bottom loop), px&2&1.79$\pm$0.10&1.79$\pm$0.12&2.13$\pm$0.12\\
\multicolumn{5}{c}{Steady loop in the vicinity of an oscillating one  (Fig. \ref{fig:model-distance}-cfi)}\\
Period (top loop), frames&20&20.0$\pm$0.1&20.0$\pm$0.1&20.0$\pm$0.1\\
Amplitude (top loop), px&2&1.76$\pm$0.11&1.64$\pm$0.11&2.07$\pm$0.08\\
Period (bottom loop), frames&-&-&19.9$\pm$0.2&20.0$\pm$0.2\\
Amplitude (bottom loop), px&0&0.00$\pm$0.00&0.13$\pm$0.11&0.27$\pm$0.03\\
\hline	
\end{tabular}
\end{table}

\section{Application to SDO/AIA data cubes}
\label{sec:observation}

\begin{figure*}
	\includegraphics[width=0.39\textwidth]{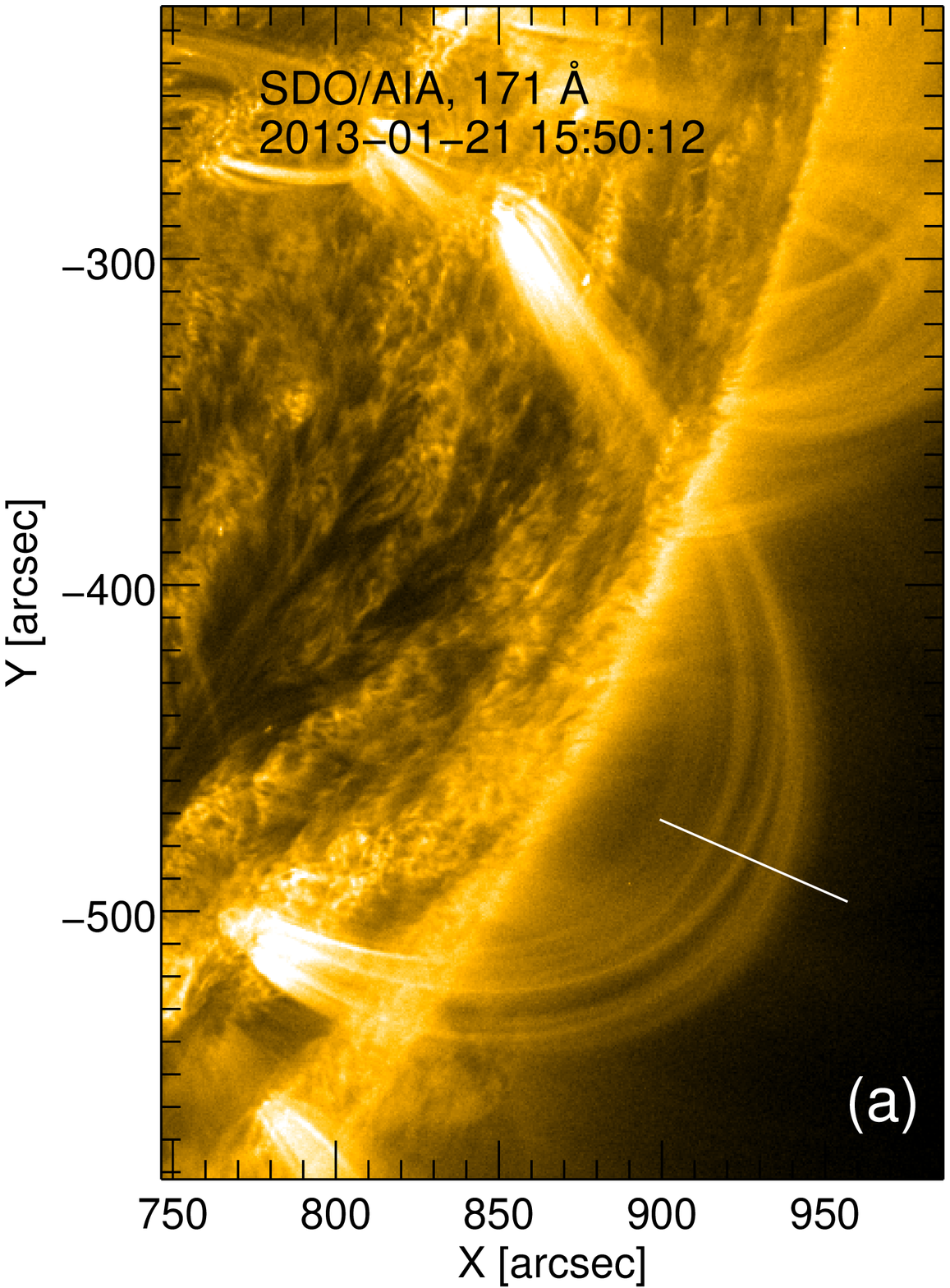}\includegraphics[width=0.61\textwidth]{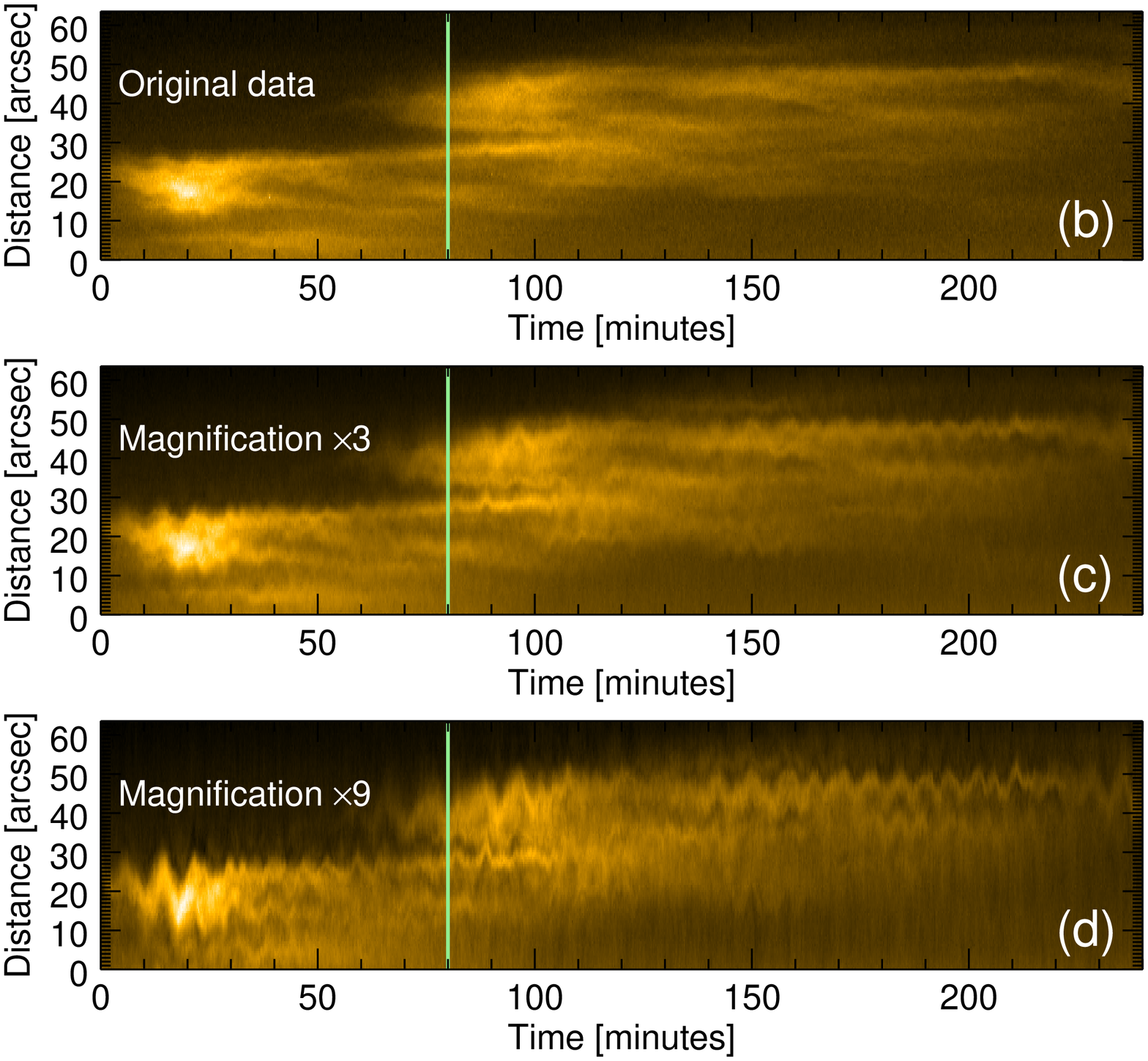} \caption{EUV image (a) of the analysed coronal loop system observed in 171~\AA~ SDO/AIA channel on 21 January 2013. Right panels show time-distance plots made with the use of the original data (b) and processed with the motion magnification technique with the magnification factors $k = 3$ (c) and $k = 9$ (d).  The vertical green line in plots (b--d) indicates the instant of time when the image (a) was taken. The artificial slit used for the time-distance plot construction is indicated by the straight white line in panel (a). The distance is measured along the slit, starting at its left edge. }
	\label{fig:real_data_decay-less} 
\end{figure*}

To test the developed method on real solar coronal data we used the SDO/AIA 171 \AA\  observations \citep{2012SoPh..275...17L} of 
two examples of the decay-less low-amplitude kink oscillations. 
The first event was previously analysed by \cite{2014A&A...570A..84N}.
 It occurred in a set of coronal loops that was observed by SDO/AIA on the south-western limb of the Sun on 21 January 2013 and was not associated with any NOAA active region. An EUV image of the loop system is shown in Fig. \ref{fig:real_data_decay-less}(a). The loop system had approximately a semi-circular shape, and consisted of many thin strands resolved with AIA. It was existing for more than 10 hours, exhibiting low-amplitude almost decay-less transverse oscillations of individual strands during its whole life time.

	As the second example, we selected  active region NOAA 11640 observed on the solar disk close to the north-western limb on 05 January 2013. This active region was previously analysed by \cite{2015A&A...583A.136A}, and was found to exhibit decay-less low amplitude kink oscillations.
	In contrast to the previous case, the emission of the loop bundle was superimposed with a varying background coming from the moss behind the loop.

\begin{figure*}
	\includegraphics[width=0.417\textwidth]{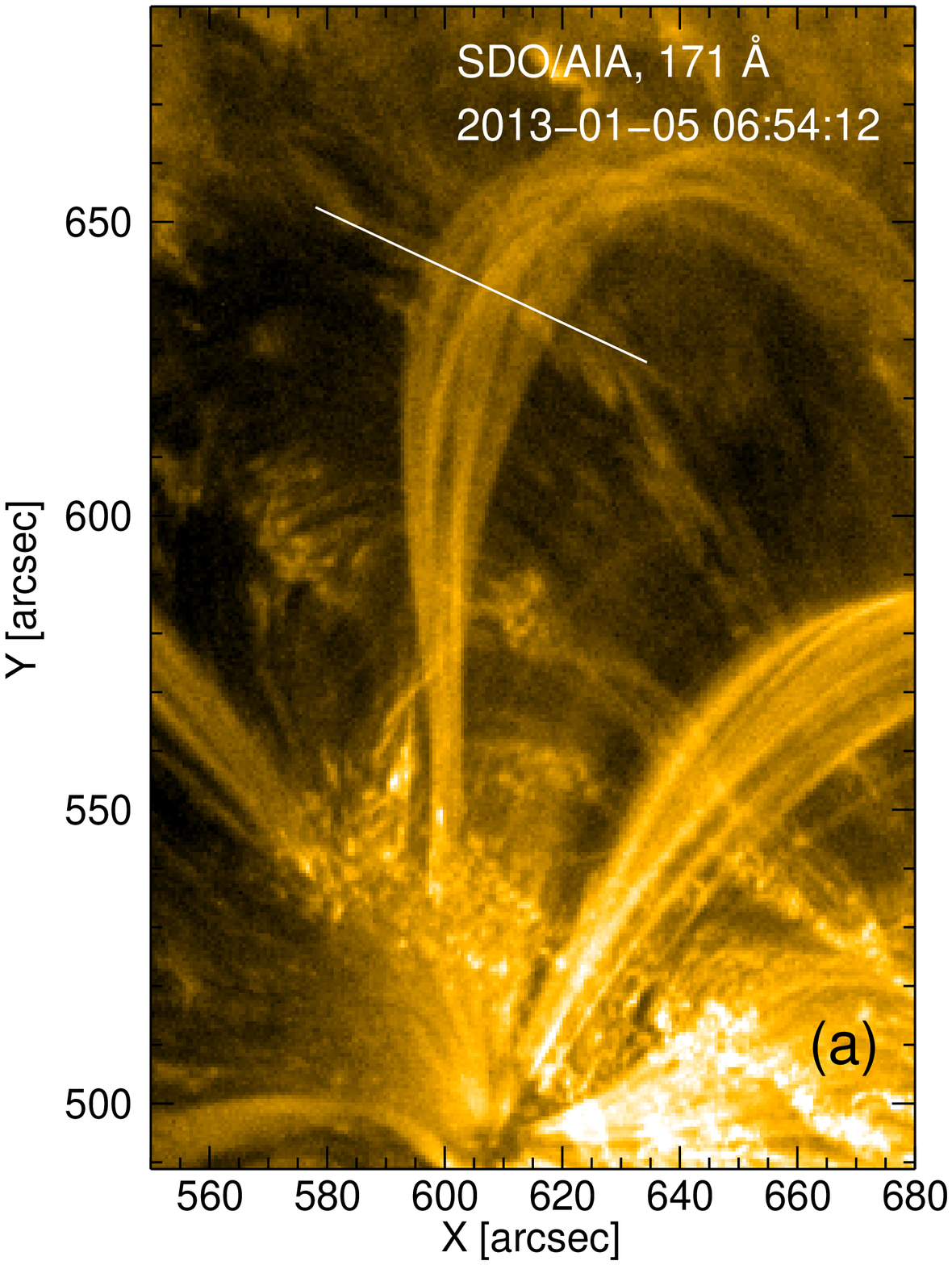}\includegraphics[width=0.583\textwidth]{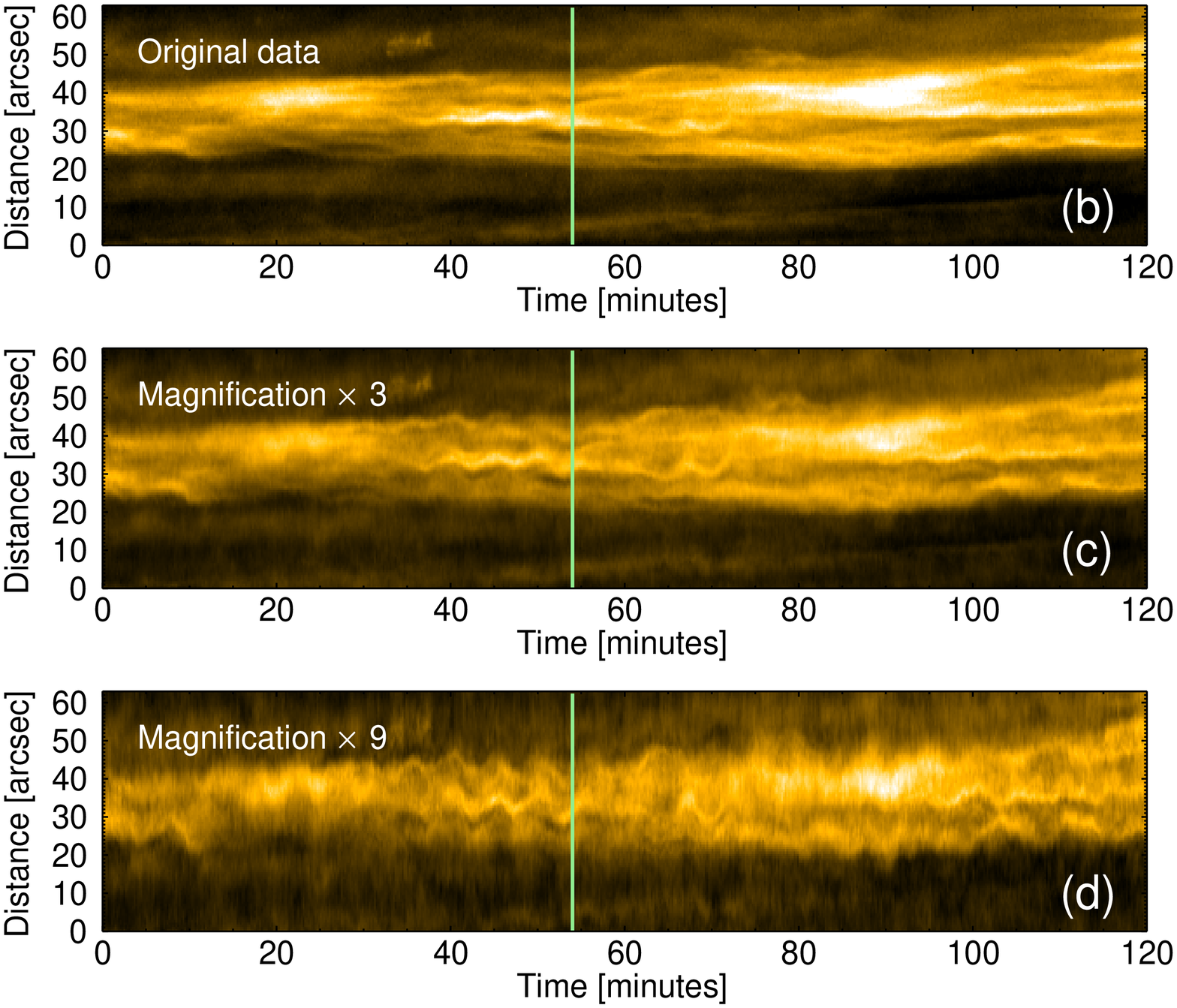} \caption{EUV image (a) of the analysed coronal loop system in active region NOAA 11640 observed in 171~\AA~ SDO/AIA channel on 05 January 2013. Right panels show time-distance plots made with the use of the original data (b) and processed with the motion magnification technique with the magnification factors $k = 3$ (c) and $k = 9$ (d).  The vertical green line in plots (b--d) indicates the instant of time when the image (a) was taken. The artificial slit used for the time-distance plot construction is indicated by the straight white line in panel (a). The distance is measured along the slit, starting at its left edge. }
	\label{fig:real_data_decay-less-disk} 
\end{figure*}

We applied the motion magnification method to both these sequences of SDO/AIA images with spatial resolution of 0.6~arcseconds and 12~s time cadence. The images were downloaded from the SDO data processing server\footnote{\url{http://jsoc.stanford.edu}}. In order to avoid the interpolation artefacts, the data was neither rescaled nor derotated. 
 The average rotation period of the Sun is about 25 days at the equator giving the linear rotation speed  of  2 km/s (or 17 SDO/AIA pixels per hour for the solar disk centre). Therefore, every object in the solar atmosphere, including coronal loops, moves across the CCD pixels of SDO/AIA with the rate of about 3.5 minutes per pixel (at the disk centre). 
 Thus, interpolations used by derotation and object tracking routines  introduce periodic changes to the data. Indeed, the simplest nearest neighbour interpolation
results in appearance of \lq\lq stairs\rq\rq-like artificial discontinuities, making incorrect even the
traditional Fourier analysis. Bi-linear or bi-cubic interpolation performs better and should
not degrade the efficiency of motion magnification. Nevertheless, we recommend not to use
interpolation before running motion magnification to exclude possible artefacts even if they are unlikely. Weak artificial periodicities  will be magnified together with the real oscillatory signal  and can become significant.

Motion magnification results
can be found in supplementary movies and 	
 are presented in Fig.~\ref{fig:real_data_decay-less}(b--d) for the off-limb observations, and in Fig.~\ref{fig:real_data_decay-less-disk}(b--d) for the on-disk observations.
As in the case of the artificial data set, the slit for the creation of time-distance plots was put  perpendicular to the loop, close to its apex.
In order to reduce noise, the slit width was set to 5 pixels, assuming that the parallel wavelength of the oscillations is much larger. Panels (b-d) of Fig.~\ref{fig:real_data_decay-less} and \ref{fig:real_data_decay-less-disk} show time-distance plots extracted both from the  original data (b) and from the motion magnification results with magnification factors of 3 (c) and 9 (d).
 
In the original time-distance plots, the low amplitude oscillations  could only be noticed by an experienced eye knowing where and when to look at. But, even a small motion magnification by a factor of 3  makes the oscillatory patterns visible to everyone without difficulty. Larger magnification ($\times 9$) allows one to investigate oscillatory profiles in detail, revealing possible amplitude evolution and coexistence of multiple harmonics.

\section{Conclusion}
\label{concl}

We developed a novel method for analysing low-amplitude transverse oscillatory motions of solar coronal plasma non-uniformities, based on the two-dimensional dual tree complex wavelet transform DT$\mathbb{C}$WT technique, the DT$\mathbb{C}$WT-based motion magnification algorithm.
The method allows us to magnify quasi-periodic transverse displacements of sub-resolution amplitude of brightness contrast structures in image sequences obtained with imaging telescopes. It makes small motions in the plane-of-the-sky to become detectable or much better visible in animations and time-distance plots. The algorithm has two parameters: the magnification coefficient $k$ and smoothing width $w$, which need to be tuned for the best performance. 

The  developed algorithm has been tested on artificial data imitating different known regimes of kink oscillations of coronal loops. It was demonstrated that the method gives good results for both harmonic and quasi-harmonic non-stationary oscillations, including exponentially decaying, frequency modulated and multi-modal oscillations. It was established that the confident detection of the oscillatory properties is possible for a broad range of the tuneable parameters $k$ and $w$, making the detection robust. The algorithm was found to give linear scaling of the magnified amplitude with the original amplitude for a broad range of small original amplitudes. Also, the magnification is independent of the oscillation period in a broad range of the periods. The technique can produce some artefacts in the case of poorly resolved, in the time domain, oscillations, when the oscillation period is shorter than 5 time frames. Also, the oscillation can \lq\lq leak\rq\rq\ to another, stationary feature in the image, situated near the oscillating one, if they are spatially close to each other. The appearance of both these artefacts can be controlled by the appropriate planning of the analysis.

We recommend to use the following analysis strategy to get reliable results with the motion magnification algorithm:
\begin{itemize}
	\item Process data cube of interest with the magnification coefficient of $\times10$ and the smoothing width corresponding to the expected period of the oscillation.
	\item Make several time-distance plots from the magnified  and original data.
	\item If there is a significant distortion in the magnified data in comparison to the original one, reduce the magnification coefficient and try again.
	\item Examine the time-distance plots and estimate the main (longest) oscillation period of interest.
	\item Set the smoothing width parameter to be slightly larger (for example, by 10~\%) than the main oscillation period and process the original data again, obtaining the final result.
\end{itemize}

The method was applied to processing of EUV observations of a non-flaring coronal active region with SDO/AIA, and allowed for a confident detection of low-amplitude decay-less kink oscillations. 

Thus, we conclude that the developed technique provides us with an effective and robust method for the study of low-amplitude kink oscillations of coronal loops, prominence threads, plumes of coronal holes, and other high contrast features. It would be especially useful for the study of the oscillation damping profiles, multi-modal oscillations, and the distribution of the oscillation amplitude along the oscillating plasma structure. In order to obtain \textbf{reasonable} results with the SDO/AIA data, we recommend to select magnification coefficient $k$ in the range from $\times$3 to $\times$10,  keeping the magnified oscillation amplitude below 10 pixels. The smoothing width $w$ should be selected slightly larger then the expected oscillation period. It is necessary to avoid considering oscillations under-sampled in the time domain, with less than 5 measurements per the expected period. Also, the analysis of oscillations of plasma structures situated close to each other in the images should be performed with caution. 

The algorithm is implemented in the Python language, and is provided in the open access online on \url{https://github.com/Sergey-Anfinogentov/motion_magnification}. The code can be transparently called from IDL, using a wrapper function provided with the code.

\begin{acks}
This work was supported by the European Research Council under the research project No. 321141 \textit{SeismoSun}. The algorithm is based on the Python code implementing DT$\mathbb{C}$WT, \url{https://github.com/rjw57/dtcwt}, developed by Dr Rich Wareham.
\end{acks}

%
%
\bibliographystyle{spr-mp-sola}
\bibliography{paper}
\end{article} 
\end{document}